# Charles Sanders Peirce and the Abduction of Einstein:
# On the Comprehensibility of the World[1]

Cornelis de Waal
cdwaal@iupui.edu

*Abstract*
Einstein was deeply puzzled by the success of natural science, and thought that we would never be able to explain it. He came to this conclusion on the ground that we cannot extract the basic laws of physics from experience using induction or deduction, and he took this to mean that they cannot be arrived at in a logical manner at all. In this paper I use Charles Peirce's logic of abduction, a third mode of reasoning different from deduction and induction, and show that it can be used to explain how laws in physics are arrived at, thereby addressing Einstein's puzzle about the incomprehensible comprehensibility of the universe. Interpreting Einstein's reflections in terms of Peirce's abduction also sheds light on abduction itself, by seeing it applied in an area where our common sense, and with that our intuitions, give us little or no guidance, and is even prone to lead us astray.

Einstein was deeply puzzled by the success of natural science and thought that we would never be able to explain it. "The eternally incomprehensible thing about the world," he famously quipped, "is its comprehensibility."[2] Unable to account for this, Einstein dubbed it a miracle (*Wunder*) and remarked that it instilled in him a "cosmic religious feeling."[3] Moreover, he added, the more we learn about the universe, the more incomprehensible it becomes that we can draw the conclusions we do.[4] Einstein's notion of comprehension is a modest one; it is merely "the production of some sort of order among sense impressions."[5]

---

[1] An earlier version of this paper was read at the *Charles S. Peirce 2014 Centennial Congress.* Lowell, Mass., July 16–19, 2014. I would like to thank the members of the audience for their valuable criticisms and suggestions.
[2] Albert Einstein, *Ideas and Opinions* (New York: Bonanza Books, 1954), 292.
[3] *Ideas and Opinions*, 39.

In part, Einstein's view that the comprehensibility of the world is an "eternal secret,"[6] is the result of his take on how natural science, especially physics, works. In this presentation I will address Einstein's puzzlement by utilizing Charles Peirce's logic of abduction. My argument runs roughly as follows. First, I discuss Einstein's views on the aim of physics and his resultant criticisms of phenomenological and empiricist approaches to physics. A consequence of this approach is that basic physical laws cannot be extracted from experience using induction or deduction, and for Einstein this means that they cannot be arrived at in a logical manner at all. Peirce, as is well known, also distinguished a third mode of reasoning, which he termed abduction or retroduction. This raises the question whether this third mode of reasoning can be used to adequately capture how in Einstein's view we arrive at these basic laws. If so, then we can establish that how these laws are being arrived at is a logical process after all, and Peirce's justification of abduction as a mode of reasoning can be used to address Einstein's puzzle as to the incomprehensible comprehensibility of the world. Interpreting Einstein's reflections in terms of Peirce's abduction also sheds light on abduction itself by seeing it applied in an area where our common sense, and with that our intuitions, give us little or no guidance and is even prone to lead us astray.

## 1. The Aim and Method of Physics

For Einstein, the aim of science is "to make the chaotic diversity of our sense-experience correspond to a logically uniform system of thought."[7] In doing this, the ideal is cover as much of our sense experiences as possible, while reducing the number of primary concepts and primary relations to a minimum. In line with this, Einstein writes in his 1933 Herbert Spencer lecture that "a complete system of theoretical physics is made up of [three things:] concepts, fundamental laws which are supposed to be valid for those concepts, and conclusions to be reached by logical deduction."[8]

---

[4] 30 March 1952 letter to Maurice Solovine; reproduced in Albert Einstein, *Letters to Solovine* (New York: Philosophical Library, 1987).
[5] *Ideas and Opinions*, 292.
[6] Ibid.
[7] *Ideas and Opinions*, 323.

This way of looking at science is reminiscent of the view Heinrich Hertz developed earlier in his *Principles of Mechanics*.[9] Hertz's description of classical mechanics may serve as an example. Hertz listed four primary concepts (space, time, force, and mass) and four primary relations, or fundamental laws, to connect them (Newton's three laws of motion and d'Alembert's principle). Once these primary concepts and relations are given, Hertz argued, everything else is just a matter of deductive inference.[10] What makes those concepts and relations primary is that they cannot be justified by the theory in which they feature; from the perspective of the theory they are purely given and non-justifiable. Hertz next presented three criteria for evaluating such axiomatic systems of primary concepts and relations: consistency, correctness, and appropriateness. The axiomatic systems should be logically permissible, not contradict observations, and in cases where multiple theories meet both demands, represent more of the relations we deem important and less relations that lack any counterpart in our observations.

With regard to theoretic structures thus conceived, Einstein argued that one should opt for theories with primary principles and concepts that are simple, few in number, and where the coordination with sense experience is both unique and convincing.[11] Moreover, the coordination should be with sense impressions *in their totality* rather than be restricted to a certain predetermined subset. Put differently, Einstein is looking for a single, global theory, not for a bunch of separate theories that are logically independent of one another.[12] A comparison of Kepler's laws with Newton's theory of gravitation can function as an example. Kepler's three laws (that the orbit of a planet is an ellipse with the Sun at one of its foci; that a line joining a planet and the sun sweeps out equal areas during equal intervals of time; and that the square of the orbital period of a planet is proportional to the cube of its mean distance to the sun) are three brilliant empirical generalizations that paint a pretty complete picture of the motions of the planets, but they are logically independent. What Newton brings to the table, in Einstein's view, is one single theory that brings the three laws together by proving them theorems of that theory, as here all physical events, including the movements of the planets, are put in terms of masses that are subject to Newton's laws of motion.[13] As Einstein observes, this move toward greater logical unification comes at a cost. The more comprehensive the framework becomes, the further its primary concepts and relations are removed from the sense experience that inspired them and forms their touchstone. This is clearly the case here, where we move from Kepler's geometrical description of the motion of the planet Mars, which applies to the other planets also, to the highly abstract ideas of force and mass that are defined contextually in terms of the theory's other primary elements. Einstein's own theory of relativity takes a step further in this direction.[14] In the end, Einstein writes, we get "a system of the greatest conceivable unity, and of the greatest poverty of concepts of the logical foundations."[15] Hence, rather than an inference to a plausible hypothesis (as Peirce does) or an inference to the best explanation (as Harman and Lipton do), the search for primary concepts and principles is best described as an inference to the greatest comprehension.

---

[8] *Ideas and Opinions*, 272.
[9] Heinrich Hertz, *Principles of Mechanics* (London: Macmillan, 1899), see esp. 1–4.
[10] Op. cit., 4–5.
[11] *Ideas and Opinions*, 323.
[12] *Ideas and Opinions*, 293.

## 2. The Quest for Primary Concepts and Primary Principles

In his discussion of mechanics, Hertz said precariously little about how we find these primary concepts and principles from which all else is supposed to follow deductively.[16] The task at hand is to specify a small number of logically independent conceptual elements in terms of which all our sense experience can be ordered; that is, to find elements that are simple and can be considered basic. For instance, in Newtonian mechanics velocity is not a logically independent concept, as it is a function of space and time. Now how do we find such basic elements? One possibility is that we extract them from experience. This is in essence the empiricist's answer. Einstein ascribes this view to Newton whom in Einstein's view "still believed that the basic concepts and laws of his system could be … deduced from experience by 'abstraction'—that is to say, by logical means."[17] For instance, we would derive the primary concept of space a posteriori by mentally removing all content from our experience, thus creating the conception of empty space. By making it a *logical* derivation, the empiricist further implies that the concepts and principles arrived at are uniquely determined (*eindeutig Bestimmt*) by our perception of the external world.

---

[13] *Ideas and Opinions*, 257.
[14] *Ideas and Opinions*, 349.
[15] *Ideas and Opinions*, 294.
[16] Admittedly, Hertz was primarily concerned with the question what justifies us in maintaining certain systems, and not in the question how its primary concepts and relations were discovered. As Hertz well knew, how the primary concepts and relations were discovered is immaterial to the system's justification.

Einstein rejected the first answer. In fact, he rejected that the process is a logical one at all, whether deductive or inductive, and he denied that the result is uniquely determined. Observing that abstraction and generalization require us to make choices, the process cannot be a logical one and, moreover, allows for alternatives. Consequently, the most we can say is that our sense impressions inspire, or hint at, certain concepts and principles. In Einstein's view, this element of choice suffices to make the resulting conceptions free creations of the human mind.[18] To this we can further add, as was said before, that when our theories become more comprehensive, like with the shift from Kepler to Newton, the primary principles and concepts become further and further removed from our sense impressions, and as a result the claim that we get them the way empiricists claim we do becomes less and less credible.

A second answer to the question of how we find these elements is that they are a product of pure reason—that they are a priori. This view goes back at least to

---

[17] *Ideas and Opinions*, 273.
[18] Albert Einstein and Leopold Infeld, *The Evolution of Physics: The Growth of Ideas From Early Concepts to Relativity and Quanta* (New York: Simon and Schuster, 1942), 33; see also *Ideas and Opinions*, 272.

Galileo who believed that keen insight—what he called *il lume naturale*—supplemented by a few observations and experiments, quickly yields them. Galileo's great successes in dynamics gave him cause to be optimistic, and writing when he did, he could still rely on a rather robust notion of God as the rational creator of the universe who in the process created us unto his own image. The situation Einstein found himself in is very different. The notion of God Galileo relied upon was no longer plausible and the role of choice in selecting primary principles and concepts had become much more pronounced, especially after the development of non-Euclidean geometries and alternative algebras. Again, the result is a view where the fundamental concepts and postulates of physics are in the logical sense free inventions of the human mind, and they are so without a plausible backstory on why or how the mind can come to comprehend the world that is external to it.[19] Not only does this mean that the primary principles and concepts are fictitious, but also that any attempt to deduce them logically from our sense experiences is, in Einstein's words, "doomed to failure."[20] That we can comprehend the world appears on this view indeed utterly incomprehensible.

## 3. Abduction

When calling a derivation logical, Einstein is purely thinking in terms of deduction and induction. Since neither applies to the process by which we obtain the primary principles and concepts, how we do obtain them can thus only be non-logical. "The scientist," he writes "has to worm [them] out of nature by perceiving in comprehensive complexes of empirical facts certain general features which permit of precise formulation."[21] For Einstein, the primary principles and concepts are "freely chosen conventions"[22] or constructions that are the product of a free play in the imagination, and as there are different ways of doing so, they are "to a large extent arbitrary."[23] Our intuition plays a central role here: "only intuition, resting on systematic understanding of experience, can reach them."[24] As such, Einstein remarks, they do not sprout from a "deliberate intention or program," but come "straight from the heart."[25] Elsewhere he describes "pure thinking" in terms of playing the violin and of smoking a pipe while lounging in an armchair.[26]

---

[19] The first principles and concepts are a priori, but not in the old sense of being necessary or analytic; they are a priori only relative to a theory.
[20] *Ideas and Opinions*, 274.
[21] *Ideas and Opinions*, 221.
[22] Albert Einstein, "Autobiographical Notes," in Paul Arthur Schilpp (ed.) *Albert Einstein: Philosopher–Scientist,* 2 vols. (New York: Harper Torchbook, 1959), 1:13.

This all being said, Einstein does give some general guides on how to go about finding them, and *The Evolution of Physics*, a book he wrote with Leopold Infeld, is in part an attempt to gain more clarity in this matter. The suggestion that there might be rules raises the question whether this decidedly non-logical process that Einstein grounded in our intuition could be recast in terms of abduction, the third mode of reasoning Peirce distinguished besides deduction and abduction, and if so where that leaves us. So let us thus turn our eye to Peirce.

For Peirce, logic is a normative science: It studies deliberate thinking with the aim of distinguishing good reasoning from bad reasoning. Peirce further added that "to say that any thinking is deliberate is to imply that it is [self-]controlled with a view to making it conform to a purpose or ideal."[27] For logic, this purpose is to attain correct representations, which is in line with Einstein's ideas about the purpose of physics discussed earlier. Peirce further maintained a naturalistic account of reasoning, arguing that it is grounded in our problem-solving activity and the problem-solving activity of many more generations than that our species is old. From this we obtained a large toolbox of practices that serve us very well in our day-to-day affairs. It is this toolbox that subsequently becomes the subject of logic, conceived as the study of how we *should* reason as opposed to how we actually do reason. Now, since reasoning developed out of our practical dealings with the world, it is far from clear how well it fares when applied to subjects that are far removed from this, as, for instance, in theoretical physics, where our common sense or intuition often fails us. Consequently, it is in such areas that the study of logic becomes of preeminent importance.

---

[23] *Ideas and Opinions*, 300.
[24] *Ideas and Opinions*, 227.
[25] Ibid.
[26] Einstein and Infeld, op. cit., 4f.
[27] Peirce Edition Project (ed.), *The Essential Peirce,* Vol. 2 (Bloomington: Indiana University Press, 1998), 376.

In its most general form, abduction can be captured as follows: Some surprising fact *B* is observed. If *A* were true, B would be explicable as a matter of course. So it is at least plausible that A is true. For instance, if we find a bag of beans and a cup of the same beans beside it, it is reasonable to assume that the beans in the cup came from the bag. Similarly, the surprising fact of the retrograde motion of Mars becomes a matter of course once we assume that Mars and Earth each orbit in an ellipse that has the sun at one of its foci. In that sense an abductive argument takes the surprise out of a surprizing experience by raising a plausible hypothesis from which the experience follows as a matter of course. As both examples suggest, abduction is here envisioned in a way that remains close to the world of sensory observations and the laws that can be abstracted from them, such as Kepler's first law, which is a favourite example of Peirce.[28] Consequently, it is not immediately clear whether abduction would apply also to the kind of situations that Einstein is talking of.

Even though Peirce explicitly admitted that abduction "is very little hampered by logical rules,"[29] he still considered it a type of reasoning. He did so in part because it is one of three ways of combining rule, case, and result (deduction and induction being the other two) and in part because it is a self-controlled deliberate process in which one tries to gain a correct or at least a plausible representation, or, to use Einstein's term, comprehension. Peirce further used the three ways of reasoning to develop a general recipe for science that resembles Einstein's portrayal of theoretical physics: we use abduction to establish an explanatory hypothesis, then use deduction to extract as many consequences of it as possible, and finally use induction to see whether these consequences match our sense experiences.

---

[28] See e.g. Cornelis de Waal (ed.), *The Illustrations of the Logic of Science by Charles S. Peirce* (Chicago, Open Court, 2014).
[29] Charles Hartshorne and Paul Weiss (eds.), *The Collected Papers of Charles Peirce,* Vol. 5 (Cambridge: Harvard University Press, 1934), sect. 188.

Before exploring whether abduction can be used to capture how Einstein's primary principles and concepts are obtained, we should see why we are so good at what in essence comes down to educated guessing. Peirce's answer is that we possess an instinct for it, and he gives an evolutionary explanation for why this is so. He writes as follows: "if the universe conforms, with any approach to accuracy, to certain highly pervasive laws or, in Einstein's words, if the universe is comprehensible], and if man's mind has been developed under the influence of those laws, it is to be expected that he should have a *natural light,* or *light of nature*, or *instinctive insight*, or genius, tending to make him guess those laws aright, or nearly aright."[30] In short, we possess reason because the world does. From Peirce's view, what makes the comprehensibility of the world incomprehensible to Einstein is that, like Galileo, Einstein makes reason into something that is independent, even alien to the world, and it is this that makes the successful application of reasoning to that world a miracle, especially when we are reasoning a priori.[31] For Peirce, in contrast, abduction is possible because our reason is the internalization, however limited and deficient, of the comprehensibility of the universe itself, and accordingly it becomes the task of logic to make this explicit with the aim of distinguishing good from bad abductions, as we are very good at the latter too. Peirce's account also suggests that abduction is more reliable when it remains close to common sense and experience, as it is within that context that we honed our talent for abduction. This also means that abduction might be less trustworthy when used for the kind of task Einstein is talking of, especially given Peirce's admission that abduction "is very little hampered by logical rules."

So far I described abduction as a self-controlled deliberate process. In Peirce's view this is not all there is to it, far from it. In common perception we find a tremendous reduction of sensory experience that follows the logic of abduction, even if this reduction is largely if not entirely unconscious.[32] So perhaps we should say that abduction, conceived as a *deliberate* process, is a special case of a non-deliberate process that is far more pervasive and that is the product of the concordant development of mind and world. It also means that a situation where we arrive at a conclusion without knowing how we got there can still be an abductive and hence a *logical* process, and this may include the process through which we find successful primary principles and concepts while playing the violin or messing with our pipe—that is, by very deliberately *not* thinking about it. Moreover, the justification of this process lies entirely in how well its products succeed, not in the method through which they were obtained.

---

[30] Op. cit., sect. 604.
[31] It is easier for Galileo to get away with this then it is for us today. Galileo could conceive of reasoning as something divine or supernatural, while maintaining that we—having been created in His image—also partake in this divine aspect, even if in a most deficient manner; the latter because our limitations as mortals and the various ways in which our non-rational elements impede and distort our reasoning.

So what, then, do we achieve by calling it a *logical* rather than a non-logical process? For one thing, the Peircean way of looking at it makes the comprehensibility of the world again comprehensible. Moreover, subjecting it to the normative science of logic makes it a process that can be inquired into, at least in principle, and thereby it avoids the charge of blocking the road of inquiry, a charge to which Einstein's view is most certainly open. Thirdly, it allows for the formation of rules, or guides, and means for evaluating processes.

One might object that by subjecting the process of discovery to logic one is putting constraints on what should be a free flow of creativity. There are at least two responses to this. First, as Peirce remarks, abduction is "very little hampered by logical rules," and second, scientists already reject many fights of the imagination. If we cannot fall back on logic, as defined by Peirce, to determine which flights of the imagination are to be taken seriously and which are not, it seems we are left with the intuitions of defenders and detractors, prematurely applied preconceived notions (including those of funding agencies), existing power relations, and the like.

Another objection, already hinted at, is that the Peircean notion of abduction remains too close to experience to be truly applicable to the type of situation

---

[32] There is even, Peirce writes in 1908, "some dose of surprise in every perception." Sheet 9 of an unpublished manuscript held in the Houghton Library at Harvard University, numbered 224 by Richard Robin in his *Annotated Catalogue of the Papers of Charles S. Peirce* (Amherst: University of Massachusetts Press, 1967).

Einstein is talking about, which is far removed from the daily observations that have shaped our instinct for guessing right. In other words, to treat the formation of hypotheses as abductive is not only too restrictive, as was first objected, but is also prone to send us in the wrong direction by suggesting that the entire universe must somehow resemble the world that we routinely experience. To this objection one may counter that if our epistemological faculty is a function of the comprehensibility of the universe, and if it this development took place not just during the age of science but during the entire history of our species and all its forebears down to its most primitive ancestors, this faculty is more likely to be reflective of the inner logic of the world than to be restricted solely to what can be surmised through our surface interactions with that world through those most remarkable abductive machines that we have come to call our senses.

    To conclude, Peirce's conception of abduction can provide us with an account of the comprehensibility of the universe, and it does so in a naturalistic way that avoids blocking the road of inquiry. The logic of abduction gives us a framework for deliberately second-guessing our guessing instinct with the aim of developing guides for the formation of new hypotheses, including those Einsteinian primary principles and concepts that give us dynamic a priori frameworks in terms of which our sense experiences could be understood.